\documentclass[aps,pra,twocolumn,twoside,superscriptaddress,notitlepage,longbibliography]{revtex4-2}

\usepackage{eufrak}
\usepackage{tikz}

\usepackage{enumitem}
\usepackage{mathtools}
\usepackage{graphicx}
\usetikzlibrary{positioning}

\usepackage{ulem} 
\usepackage{lineno}
\usepackage[caption=false]{subfig}
\usepackage{dcolumn}
\usepackage{amsmath,amssymb}
\usepackage{bm}
\usepackage{bbm}
\usepackage{overpic}
\usepackage{latexsym}
\usepackage{color}
\usepackage[english]{babel}
\usepackage{latexsym}
\usepackage{psfrag,graphicx}
\usepackage{epsf}
\usepackage{amsmath}
\usepackage{amssymb}
\usepackage{amsfonts}
\usepackage{natbib}

\usepackage{appendix}
\usepackage{verbatim}
\usepackage{enumitem}
\usepackage{dsfont}

\usepackage{float}
\usepackage{tikz}
\makeatletter
\def\grd@save@target#1{%
	\def\grd@target{#1}}
\def\grd@save@start#1{%
	\def\grd@start{#1}}
\tikzset{
	grid with coordinates/.style={
		to path={%
			\pgfextra{%
				\edef\grd@@target{(\tikztotarget)}%
				\tikz@scan@one@point\grd@save@target\grd@@target\relax
				\edef\grd@@start{(\tikztostart)}%
				\tikz@scan@one@point\grd@save@start\grd@@start\relax
				\draw[minor help lines,magenta] (\tikztostart) grid (\tikztotarget);
				\draw[major help lines] (\tikztostart) grid (\tikztotarget);
				\grd@start
				\pgfmathsetmacro{\grd@xa}{\the\pgf@x/1cm}
				\pgfmathsetmacro{\grd@ya}{\the\pgf@y/1cm}
				\grd@target
				\pgfmathsetmacro{\grd@xb}{\the\pgf@x/1cm}
				\pgfmathsetmacro{\grd@yb}{\the\pgf@y/1cm}
				\pgfmathsetmacro{\grd@xc}{\grd@xa + \pgfkeysvalueof{/tikz/grid with coordinates/major step}}
				\pgfmathsetmacro{\grd@yc}{\grd@ya + \pgfkeysvalueof{/tikz/grid with coordinates/major step}}
				\foreach \x in {\grd@xa,\grd@xc,...,\grd@xb}
				\node[anchor=north] at (\x,\grd@ya) {\pgfmathprintnumber{\x}};
				\foreach \y in {\grd@ya,\grd@yc,...,\grd@yb}
				\node[anchor=east] at (\grd@xa,\y) {\pgfmathprintnumber{\y}};
			}
		}
	},
	minor help lines/.style={
		help lines,
		step=\pgfkeysvalueof{/tikz/grid with coordinates/minor step}
	},
	major help lines/.style={
		help lines,
		line width=\pgfkeysvalueof{/tikz/grid with coordinates/major line width},
		step=\pgfkeysvalueof{/tikz/grid with coordinates/major step}
	},
	grid with coordinates/.cd,
	minor step/.initial=.2,
	major step/.initial=1,
	major line width/.initial=2pt,
}
\makeatother

\makeatletter
\def\resetMathstrut@{%
	\setbox\z@\hbox{%
		\mathchardef\@tempa\mathcode`\[\relax
		\def\@tempb##1"##2##3{\the\textfont"##3\char"}%
		\expandafter\@tempb\meaning\@tempa \relax
	}%
	\ht\Mathstrutbox@\ht\z@ \dp\Mathstrutbox@\dp\z@}
\makeatother
\begingroup
\catcode`(\active \xdef({\left\string(}
\catcode`)\active \xdef){\right\string)}
\endgroup
\mathcode`(="8000 \mathcode`)="8000

\definecolor{mygrey}{gray}{0.35}
\definecolor{myblue}{rgb}{0.2,0.2,0.8}
\definecolor{myzard}{cmyk}{0,0,0.05,0}
\definecolor{mywhite}{rgb}{1,1,1}
\definecolor{myred}{rgb}{0.9,0.1,0.}
\usepackage[colorlinks=true,citecolor=myblue,linkcolor=myblue,urlcolor=myblue]{hyperref}

\usepackage[makeroom]{cancel}

\newtheorem{theorem}{Theorem}

\newtheorem{proposition}[theorem]{Proposition}

\newenvironment{proof-of}[1]{\medskip\noindent\textbf{Proof of {#1}.}}{\hfill$\blacksquare$\medskip}

\newcommand{\tr}{\operatorname{\bf{tr}}} 

\newcommand{\diag}{\operatorname{\bf{diag}}} 
\newcommand{\ket}[1]{\vert #1 \rangle} 
\newcommand{\bra}[1]{\langle #1 \vert} 

\renewcommand{\vec}[1]{\text{\boldmath$#1$}}

\newcommand{\norm}[1]{\ensuremath{\left|\left|#1\right|\right|}}

\newcommand{\braket}[2]{\ensuremath{\langle #1 | #2 \rangle}}

\newcommand{\abs}[1]{\left\lvert #1\right\rvert}

\newcommand{\im}{\mathrm{Im}}
\newcommand{\re}{\mathrm{Re}}

\newcommand{\be}{\begin{equation}} 							
\newcommand{\ee}{\end{equation}}
\newcommand{\bea}{\begin{eqnarray}}
\newcommand{\eea}{\end{eqnarray}}
\newcommand{\bematrix}{\left(\begin{matrix}}
\newcommand{\ematrix}{\end{matrix}\right)}
\def\one{{\mbox{$1 \hspace{-1.0mm}  {\bf l}$}}}

\def\tr{\mathrm{Tr}}

\usepackage{tcolorbox}

\newcommand{\gs}[1]{{\color{black}#1}}

\begin{document}

\title[ATI]{Volumes of parent Hamiltonians for benchmarking quantum simulators}

\author{Mar\'ia Garc\'ia D\'iaz}
\affiliation{F\'{\i}sica Te\`{o}rica: Informaci\'{o} i Fen\`{o}mens Qu\`{a}ntics, %
	Departament de F\'{\i}sica, Universitat Aut\`{o}noma de Barcelona, 08193 Bellaterra, Spain}

\author{Gael Sent\'is}
\affiliation{F\'{\i}sica Te\`{o}rica: Informaci\'{o} i Fen\`{o}mens Qu\`{a}ntics, %
	Departament de F\'{\i}sica, Universitat Aut\`{o}noma de Barcelona, 08193 Bellaterra, Spain}

\author{Ramon Mu\~noz Tapia}
\affiliation{F\'{\i}sica Te\`{o}rica: Informaci\'{o} i Fen\`{o}mens Qu\`{a}ntics, %
	Departament de F\'{\i}sica, Universitat Aut\`{o}noma de Barcelona, 08193 Bellaterra, Spain}

\author{Anna Sanpera}
\affiliation{F\'{\i}sica Te\`{o}rica: Informaci\'{o} i Fen\`{o}mens Qu\`{a}ntics, %
	Departament de F\'{\i}sica, Universitat Aut\`{o}noma de Barcelona, 08193 Bellaterra, Spain}
\affiliation{ICREA, Pg. Llu\'is Companys 23, 08010 Barcelona, Spain}

\date{\today}

\begin{abstract}
 We investigate the relative volume of parent Hamiltonians having a target ground state up to some fixed error $\epsilon$, a quantity which sets a benchmark on the performance of quantum simulators.
For vanishing error, this relative volume is of measure zero, whereas for a generic $\epsilon$ we show that it increases with the dimension of the Hilbert space. 
We also address  
the volume of parent Hamiltonians when they are restricted to be local. For translationally invariant  Hamiltonians, we provide an upper bound to their relative volume. 
Finally, we estimate numerically the relative volume of parent Hamiltonians when the target state is the ground state of the Ising chain in a transverse field.  
\end{abstract}
\maketitle
Quantum simulators aim at implementing non-trivial many-body Hamiltonians whose ground state, low energy physics or dynamics are not well understood. 
The interactions embedded in such Hamiltonians give rise to highly complex quantum correlations, making analytical or numerical solutions in general unfeasible. 
Often, however, the problem of interest is the inverse one: 
given a specific relevant many-body ground state, 
which are the parent Hamiltonians that generate it?

Generic properties of Hamiltonians without a prior knowledge of their explicit form can be derived from a measure theoretical approach, as shown 
in random matrix theory to study level repulsion \cite{dyson1962}, transport phenomena \cite{beenakker1997} or atomic spectra of complex atoms \cite{Frisch2014}. 
Also, such technique was 
employed to analyze storage capacities of attractor neural networks~\cite{Gardner1988}.

We use a measure theoretical approach to calculate the probability that by randomly sampling a Hamiltonian one obtains
the parent  Hamiltonian of a targeted ground state. That is, for a given set of Hamiltonians with some specifications (e.g., dimensions, symmetries, number of parties, locality, etc.), we 
estimate the proportion of them that have a ground state which is sufficiently close to the target one. For general Hamiltonians with the only restriction of constant dimensionality, this quantity can be viewed as the probability that a given quantum state of that dimension appears as a physically meaningful state. Furthermore, for a universal quantum simulator, such probability provides  \gs{one possible} benchmark on its minimal performance at implementing quantum states, i.e., 
 it tells how likely a target quantum state can be  sufficiently well approximated.

Our problem
bears resemblances with estimating the volume of quantum states \cite{Zyczkowski1998,bengtsson2006}, 
the volume of quantum maps realizing a given task \cite{Szarek2008}, or the volume of their corresponding Choi states \cite{Lewenstein2020}. What is different here is that Hamiltonians  present a richer internal structure arising, for instance, from locality constraints or frustration \cite{davidPG2015}. Moreover, the diagonalization of a Hamiltonian provides eigenstates which are endowed with a physical meaning, while for a quantum state it yields one of the possible equivalent ensembles that realizes it. 

 A particularly relevant set of many-body Hamiltonians are those whose interactions take place between a restricted number of parties. Such local structure of the Hamiltonian has profound implications on the entanglement and correlations of their corresponding eigenstates. Finding the ground state of such Hamiltonians, the so-called local Hamiltonian problem, is NP-hard \cite{kitaev2002,kempe2006}. 
 The analysis of the volume of local parent Hamiltonians, which is dual to the volume of such special ground states, thus provides  a novel perspective on the local Hamiltonian problem.

Before proceeding further, let us summarize  our main results. Here we restrict ourselves to non-degenerate bounded Hamiltonians in arbitrary finite dimension. Despite the physical relevance of gapless Hamiltonians, its volume is of measure zero in the manifold of Hamiltonians. Under such premises, we first show
that the relative volume of parent Hamiltonians with an exact target ground state is of measure zero. When allowing for some deviation from  the target ground state, though, this volume is finite and increases with the dimension of the Hilbert space. This implies that implementing a ground state up to some fixed tolerance is more likely in higher-dimensional spaces than in lower-dimensional ones. We then address the problem of computing the relative volume of local Hamiltonians.  
The locality restriction renders the problem far more difficult. Nevertheless, we provide an upper bound for the specific case of $t$-local translationally invariant (TI) Hamiltonians. Finally, we numerically tackle the relative volume 
for the ground state of the quantum transverse Ising model, and 
compute how many 2-local non-translationally invariant Hamiltonians are parent to it up to some fidelity.
For ease of exposition, we defer the proofs of Theorems and Propositions to the Appendix. \\

\gs{\section{Volume of the manifold of Hamiltonians}}
 Let ${\cal{H}}_N$ be an $N$-dimensional Hilbert space and  $\mathcal{B}({\cal{H}}_N)$  the set of  its bounded operators. 
We denote by $\mathbf{H}_{N,k}:=\{H\in \mathcal{B}({\cal{H}}_N) : H>0;  \tr H \leq k\}$, the manifold of
positively defined Hamiltonians with trace equal or smaller than $k> 0$. Since any 
non-positive definite Hamiltonian $H'$ can always be transformed into a positive one by freely shifting up its 
eigenenergies, 
it suffices to calculate the volume of  $\mathbf{H}_{N,k}$ for $k$ sufficiently large.

\gs{Any $H\in \mathbf{H}_{N,k}$ can be expressed as  $H=UDU^\dagger$, where $D=\text{diag}(\lambda_1,\lambda_2,...,\lambda_{N})$,  with $\lambda_i > 0 \; \forall i$, $\lambda_i\neq \lambda_j \; \forall i,j$,   $ \tr H=\sum_i\lambda_{i}\leq k$,  and $U$ is a unitary matrix. 
The volume of this (convex) manifold can be computed with respect to several {\it bona fide} metrics, such as the ones induced by the Hilbert-Schmidt (HS), the Bures or the trace distance \cite{bengtsson2006}. The main results of our work do not depend on the choice of the metric, as shown later. 
Here we choose the measure generated by the Hilbert-Schmidt (HS) distance,
$d_{\text{HS}}(A,B)=\sqrt{\tr (A-B)^2}$, for two Hermitian operators $A$ and $B$, inasmuch as the HS distance is simpler to deal with, it induces the Euclidean geometry into the manifold of Hermitian operators \cite{Zyczkowski2003,wang2009}, and it is widely used in quantum information tasks \cite{ozawa2000,lee2003,Pandya2020,Arrasmith2019}.

The first step to estimate the volume of the manifold is to obtain the infinitesimal distance $d_{\text{HS}}(H,H+dH)$, giving  
rise to its line element  $ds^2:=d^2_{\text{HS}}(H,H+dH)=\tr [(dH)^2]$, where $dH= U(dD+U^\dagger dU D-DU^\dagger dU)U^\dagger$, leading to
\begin{eqnarray}\label{ds_general}
ds^2&=&\sum_{i=1}^{N}(d\lambda_i)^2 
+2\sum_{i<j}^N(\lambda_i-\lambda_j)^2|(U^\dagger dU)_{ij}|^2,
\end{eqnarray}
where we have used $UdU^\dag=-dU U^\dag$ (see \cite{Zyczkowski2003} for more details).
Notice that the two sets of variables $\{d\lambda_i\}$ and $\{\re (U^\dagger dU)_{ij},\im (U^\dagger dU)_{ij}\}$ do not get mixed up in the line element, yielding a block-diagonal metric tensor whose determinant (in absolute value) corresponds to the squared magnitude of the Jacobian determinant of the transformation $H\rightarrow UDU^\dagger$. Hence, the volume element of the manifold reduces to the product form $dV=d\mu(\lambda_1,...,\lambda_{N})\times d\nu_{\text{Haar}}$, where the first factor depends only on the eigenvalues of $H$ and the second one corresponds to the Haar measure on the $N$-dimensional complex flag manifold $ Fl_{\mathbb{C}}^{(N)}:=U(N)/[U(1)^N]$, where $U(N)$ denotes the unitary group in dimension $N$. Indeed, a volume element of the referred form is specific to all unitarily-invariant measures, since the Haar measure is unitarily-invariant. After integration we arrive to}\\
%



\begin{proposition}\label{volume_all}
The HS volume of the manifold $\mathbf{H}_{N,k}$ 
amounts to
\begin{equation}
\label{vol-N}   
\normalfont{\text{vol}_{N}(\mathbf{H}_{N,k})}=I_1(N,k)I_2(N),
\end{equation}
where 
\begin{eqnarray}
I_1(N,k)=\frac{\sqrt{N}}{N^2! \; N!} \xi_N \xi_{N-1} k^{N^2},
\end{eqnarray}
with $\xi_n=\Pi_{j=1}^n j!$, comes from the integration over the simplex of eigenvalues, and
\begin{eqnarray}\label{vol_flag}
I_2(N)= {\rm vol}_N( Fl_{\mathbb{C}}^{(N)})=\frac{(2\pi)^{N(N-1)/2}}{\xi_{N-1}}
\end{eqnarray}
corresponds to the Haar volume of the unitaries over the complex flag manifold. 
\end{proposition}

In passing we stress that the volume of the set of density matrices \cite{Zyczkowski2003}, $\rho\in {\cal B}({\cal{H}}_N)$ s.t. $\rho\geq 0$ with $\tr(\rho)=1$, is the boundary surface of Eq.~\eqref{vol-N} for $k=1$, i.e., $\partial_k \text{vol}_{N}(\mathbf{H}_{N,k})|_{k=1}$.\\

\gs{\section{Relative volume of Hamiltonians with a target ground state}}
The relative volume gives the probability of randomly sampling a Hamiltonian $H\in \mathbf{H}_{N,k}$ that is parent to a target state $\ket{\psi_0}$. Such Hamiltonians constitute the manifold $\mathbf{H}_{N,k}^{\ket{\psi_0}}\subset \mathbf{H}_{N,k}$, whose volume
results from integrating over all unitaries in $U(N-1)$.  
Since the volume of a manifold is basis-independent, one can always choose a basis where $\ket{\psi_0}:=\ket{0}=(1,0,...,0)^T$.
As the columns of $U$ have to form an orthonormal basis, it follows that $U= 1 \oplus U'$, 
where $U'\in Fl_{\mathbb{C}}^{(N-1)}$ (recall that $U$ is uniquely specified if it belongs to the complex flag manifold). Thus, integrating over $U'$ leads to a volume in one dimension less, that is, an \mbox{$(N-1)$}-dimensional hypersurface of $\mathbf{H}_{N,k}$:
\begin{proposition}\label{vol_fixedground}
	The HS hypersurface of the manifold $\mathbf{H}_{N,k}$ 
	with a target ground state $\ket{\psi_0}$ is given by
\begin{equation}\label{vol_hypersurface}
	S^{(1)}_{N}(\mathbf{H}_{N,k}^{\ket{\psi_0}})=I_1(N,k) I_2(N-1).
\end{equation}
\end{proposition}
Accordingly, the volume of $N$-dimensional Hamiltonians with $L$ fixed eigenstates is actually a hypersurface
$S_{N}^{(L)}(\mathbf{H}_{N,k}^{\ket{\psi_0},...,\ket{\psi_{L-1}}})=I_1(N,k)I_2(N-L)$.
%
By construction, the hypersurface of (unrestricted) Hamiltonians with a specified ground state does not depend on the choice of the latter. This will not be the case when imposing further structure on $\mathbf{H}_{N,k}^{\ket{\psi_0}}$, e.g., when considering \emph{local} Hamiltonians. For instance, Matrix Product State (MPS) ground states are unique ground states of local, gapped, frustration free Hamiltonians \cite{davidPG2015}.

%

Still, Eq.~\eqref{vol_hypersurface} is an absolute volume, and as such tells little 
about the relative occurrence of Hamiltonians with a common ground state in a given dimension. Instead, the relative volume
$\text{vol}_r(\mathbf{H}_{N,k}^{\ket{\psi_0}}):=\text{vol}_{N}(\mathbf{H}_{N,k}^{\ket{\psi_0}})/
\text{vol}_{N}(\mathbf{H}_{N,k})$ 
 is the meaningful quantity. However, $\mathbf{H}_{N,k}^{\ket{\psi_0}}$ and $\mathbf{H}_{N,k}$  
 refer to manifolds of different, and thus incomparable, dimensions. We address this issue by tolerating a small deviation $\epsilon$ from the ground state $\ket{\psi_0}$, that is, we consider the volume of the manifold $\mathbf{H}_{N,k}^{\ket{\psi_0^\epsilon}}$, corresponding to Hamiltonians with ground states $\ket{\psi_0^\epsilon}$ such that $|\braket{\psi_0}{\psi_0^\epsilon}|\geq 1-\epsilon$. In doing so, we extend the hypersurface Eq.~\eqref{vol_hypersurface} into a volume in $N$ dimensions which is directly comparable with Eq.~\eqref{vol-N}, enabling a proper definition of a relative volume: 
\begin{proposition}\label{vol_fixed_eps}
The HS volume of the manifold $\mathbf{H}_{N,k}^{\ket{\psi_0^\epsilon}}$ 
with ground state $\ket{\psi_0^\epsilon}$ such that $|\braket{\psi_0}{\psi_0^\epsilon}|\geq 1-\epsilon$ for sufficiently small $\epsilon$ is given by
\begin{align}
\normalfont{\text{vol}}_{N}(\mathbf{H}_{N,k}^{\ket{\psi_0^\epsilon}})
	&= {I_1(N,k)\int_{Fl_{\mathbb{C}}^{(N)}} \mathds{1}_{[1-\epsilon,1]}\left(\abs{\!\bra{\psi_0}U\ket{0}\!}\right) \nonumber}\\
&{\times |\prod_{i<j} 2\re(U^\dagger dU)_{ij}\im(U^\dagger dU)_{ij}| \nonumber}\\
	&\approx \epsilon I_1(N,k) I_2(N-1),
\end{align}
where  $\mathds{1}_{[1-\epsilon,1]}$ is the indicator function.
\end{proposition}

One immediately obtains:

\begin{proposition}\label{vol_rel_prop}
The relative volume of $\mathbf{H}_{N,k}^{\ket{\psi_0^\epsilon}}$ is
\begin{equation}
\normalfont{\text{vol}}_r(\mathbf{H}_{N,k}^{\ket{\psi_0^\epsilon}})\approx \epsilon \frac{I_2(N-1)}{I_2(N)}\approx \epsilon (2\pi)^{1-N}(N-1)!. 
\end{equation}
\end{proposition}

\begin{figure}
	\centering
	\includegraphics[width=1\linewidth]{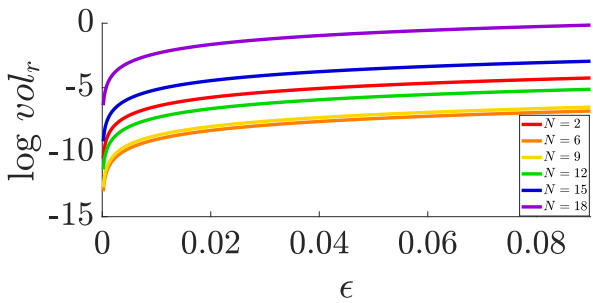}
	\caption{Logarithm of the relative volume \gs{of} the manifold $\mathbf{H}_{N,k}^{\ket{\psi_0^\epsilon}}$ as a function of the error $\epsilon$ for different dimensions of the Hilbert space $N$.} %
	\label{fig:volR_eps_c}
\end{figure}

As expected, this probability vanishes for $\epsilon=0$, meaning that the subset of Hamiltonians with an exact target ground state is of measure zero.  Fig. \ref{fig:volR_eps_c} shows the behaviour of the relative volume in logarithmic scale with respect to $\epsilon$ 
as the Hilbert space dimension increases. The relative volume monotonically increases with the dimension of $\mathcal H$
except for lower values of $N$ where the monotonicity is lost due to the singular behaviour already shown by the volume of the unitary ball in small dimensions. For sufficiently large $N$, a better insight can be obtained by using Stirling's formula leading to $\normalfont{\text{vol}}_r(\mathbf{H}_{N,k}^{\ket{\psi_0^\epsilon}})\approx \epsilon (2\pi/e)^{-N} \, N^{N}$. Clearly, the relative volume increases with $N$, for $N\gg 1$. However, as the relative volume should be smaller than one, this imposes as well a restriction in the maximal compatible error,
$\epsilon\lesssim (2\pi/e)^{N} \, N^{-N}$. Thus, for large $N$, and as long as this last inequality holds, a higher-dimensional Hamiltonian is more likely to be  parent to a target ground state up to some fixed 
error than the corresponding lower-dimensional one.

The above results can be easily adapted to real Hamiltonians by properly modifying the line element of the manifold to the real case in  Eq.~(\ref{ds_general}). For such a case, 
$\text{vol}_r^{(\mathbb{R})}(\mathbf{H}_{N,k}^{\ket{\psi_0^\epsilon}})\approx \epsilon \;  2^\frac{(1-N)}{2}\pi^{-\frac{N}{2}}\Gamma(\frac{N}{2})$, \gs{with $\Gamma$ the Euler gamma function \cite{abramowitz}}, as explicitly shown in Appendix \ref{appendix1}. Roughly speaking,  this result reflects the fact  that the number of free parameters in real Hamiltonians is reduced by half.

Finally, we note that although the volume of the manifold of Hamiltonians depends on the used metric, the relative volume does not, as far as the measure is unitarily invariant. Indeed, the dependence on the metric in  $\text{vol}_{N}(\mathbf{H}_{N,k})=I_1(N,k)I_2(N)$
appears in the term $I_1(N,k)$, which is a function of the eigenvalues of $H$.  The use of another metric will lead to a different $\tilde{I}_1(N,k)$. As relative volumes do not depend on this term, all unitarily-invariant measures yield the same  relative volume.\\

\gs{\section{Volume of local Hamiltonians}}
Physically relevant Hamiltonians are usually local. 
An $n$-body $t$-local Hamiltonian is of the form $H=\sum_{i=1}^{M}h_i$, where $h_i$ is a Hamiltonian acting non-trivially on at most  $t$ parties, and $M$ is some positive integer. 
Such $t$-local Hamiltonian  can be viewed as a set of $M$ constraints on the $n$ parties, each involving at most $t$ of them.

A way to calculate the volume of such manifold 
amounts to diagonalizing each of the $M$ $t$-local Hamiltonians,  $h_i=u_i \Lambda_i u_i^\dagger$, where $\Lambda_i$ are  diagonal matrices of eigenvalues, and $u_i$ are the corresponding unitary matrices. Defining $dG_i:=u_i^\dagger du_i$, the line element of this manifold becomes  
\begin{eqnarray}\label{metric_local}
ds^2&=&
\sum_{i=1}^{M}(\sum_{k=1}^{N}(d\Lambda_{ik})^2 +\sum_{k\neq l}^{N}(\Lambda_{ik}-\Lambda_{il})^2|(dG_i)_{kl}|^2) \nonumber \\
&+& \sum_{i\neq j}^M\tr \left(u_i(d\Lambda_i+dG_i \Lambda_i -\Lambda_i dG_i)u_i^\dagger \right.\nonumber\\
&\times& \left. u_j( d\Lambda_j+dG_j \Lambda_j 
-\Lambda_j dG_j)u_j^\dagger \right).
\end{eqnarray}
Although the first term of the line element can be treated in the same manner as in Eq.~(\ref{ds_general}), the second one contains crossed terms $dh_idh_j$ which turn out to be an involved function of the eigenvalues and eigenvectors both of $h_i$ and $h_j$, preventing to obtain an evaluable expression for the volume of local Hamiltonians. 

Let us restrict to translationally invariant (TI) Hamiltonians, i.e., those of the form $H=\sum h_i$ where all $h_i$ are locally equal, that is,
$h_i\equiv \mathds{1}\otimes\cdots\otimes\mathds{1}\otimes h^{(i)}\otimes\mathds{1}\otimes\cdots\otimes\mathds{1}$, where   
$h$ is a $d^t$-dimensional Hamiltonian acting on $t$ $d$-dimensional parties, and the multi-index $i$ labels the set where $t$ acts.  If we restrict to 1D models, the multi-index $i$ refers to the first particle in which $h$ acts.

Even in the simplest instance of TI ($n=3$, $d=t=2$) the calculation of the volume remains out of reach. 
Take $H=h_1+h_2=h\otimes \mathds{1}+\mathds{1}\otimes h$ with $\dim(h)=2^2$. 
The line element of this manifold reads 
\begin{eqnarray}
ds^2&=&4\sum_{k=1}^4(d\Lambda_k)^2 +4\sum_{k\neq l}^4(\Lambda_k-\Lambda_l)^2|dG_{kl}|^2 \nonumber \\
&+& 2\tr \left(u(d\Lambda+dG \Lambda -\Lambda dG)u^\dagger \right.\nonumber\\
&\times& \left. Pu( d\Lambda+dG \Lambda 
-\Lambda dG)u^\dagger P^\dagger \right),
\end{eqnarray}
where $h_1=u \Lambda u^\dagger$, $h_2=(Pu)\Lambda (Pu)^\dagger$ and $dG=u^\dagger du$, with $u$ a unitary matrix, $\Lambda$ a diagonal matrix of eigenvalues and $P$ a permutation of every row of $u$ except for the first and last ones. 
Even though the second term only depends on a single unitary $u$ and a permutation matrix $P$, the metric still involves hundreds of terms 
\footnote{Approximating $u^\dagger P u\simeq \one$ is simply wrong, notice that for a general $N$-dimensional unitary matrix $U$ it holds  $\norm{\mathds{1}-U^\dagger P U}_{\text{HS}}=\norm{U^\dagger(\mathds{1}- P)U}_{\text{HS}}=\norm{\mathds{1}- P }_{\text{HS}}=\sqrt{N-2}\gg 0$.}. Nevertheless, we find an upper bound to the volume of TI Hamiltonians by considering that the local terms $h_i$ are equal but act in disjoint subspaces, i.e., $H=\bigoplus_i h_i$. For such Hamiltonians, the line element would be given by the (corresponding) Eq.~\eqref{ds_general}, permitting the computation of the volume. 
Locally non-overlapping TI Hamiltonians are subject to fewer constraints 
than their generic  TI counterparts, and thus the volume of the former should upper bound that of the latter. 
To see why, let us express the Hamiltonian in terms of the generators of the corresponding algebra. Any  $t$-local TI Hamiltonian can be expressed as $H=\sum_{l=1}^M h_l$, with  
$h_l\equiv h=\sum_{i,j,...,k=0}^{d^2-1}\alpha_{ij...k}\underbrace{\sigma_i \otimes \sigma_j \otimes ... \otimes\sigma_k}_{t},\; \forall l$.
The set $\{\sigma_m \}$ denotes the generators of $SU(d)$ and the identity, forming a proper basis of $\mathcal{B}(\cal{H})$. The coefficients $\alpha_{i j...k}$ are real  and independent. Suppose that such a manifold is associated to some metric tensor $g$. Now, removing the crossed terms $dh_idh_j$ 
from the line element in Eq.~\eqref{metric_local} results in a diagonal metric tensor $\tilde{g}$, such that $g=\tilde{g}+X$,  where $X$ is a matrix with vanishing diagonal. Due to the positivity of metric tensors, Hadamard's inequality \cite{rozanski2017} can be applied to show  that $\det(g)\leq \det (\tilde{g})$. Therefore, calculating the volume associated to the line element without crossed terms yields an upper bound for the volume of the manifold of $t$-local TI Hamiltonians of dimension $N$ ($\mathbf{H}_{N,k,t}^{\rm TI}$), as formally demonstrated in Appendix \ref{appendix2}.\\

%
\begin{theorem}\label{bounds}
The HS volume of the $t$-local manifold 
 $\mathbf{H}_{N,k,t}^{\rm TI}$ 
is upper bounded by: 
\begin{eqnarray}\label{bound_eq}
 {\rm vol}_{d^t}(\mathbf{H}_{N,k,t}^{\rm TI})
 \leq \nu^{\frac{\kappa}{2}}  I_1(d^t,\frac{k}{\nu})I_2(d^t),
\end{eqnarray}
where $\nu=M d^{n-t}$ and $\kappa=d^{2t}-1$.
\end{theorem}

Like the absolute volume of generic Hamiltonians [Eq.~\eqref{vol-N}], this bound decreases with increasing number of parties $n$. However, the prefactor $\nu^{\frac{\kappa}{2}}$ makes the bound increase with the local dimension $d$ and with the locality factor $t$.


The $d^t$-dimensional volume  in Eq.~\eqref{bound_eq}  upper bounding the volume of $t$-local TI Hamiltonians is of measure zero with respect to the $d^n$-dimensional volume of all Hamiltonians with the same number of parties. Thus, an upper bound for the relative volume cannot be defined under the TI restriction.  To shed some light on this question, we now allow for locality to be broken up to a small extent. Consider a Hamiltonian of the form $H=h_{\text{TI}}+\delta h_{\text{NL}}$, where $h_{\text{TI}}$ is a TI Hamiltonian, $h_{\text{NL}}$ is a generic nonlocal Hamiltonian, and $\delta \ll 1$.
Embedding the manifold of $t$-local TI Hamiltonians in such a $d^n$-dimensional manifold now permits the definition of a relative volume. Applying the same arguments leading to Theorem \ref{bounds}, one obtains 
\begin{theorem}
The relative volume of $d^n$-dimensional $\delta$-TI Hamiltonians  $H=h_{\text{TI}}+\delta h_{\text{NL}}$, with $h_{\text{TI}}$  a $t$-local TI Hamiltonian such that $\tr h_{\text{TI}}\leq k$, $\delta \ll 1$ and $h_{\text{NL}}$  a general nonlocal Hamiltonian with $\tr h_{\text{NL}} \leq k'\leq k$, fulfills

\begin{equation}
\normalfont{\text{vol}}_r(\mathbf{H}_{N,k,k',t}^{\delta\text{-TI}}) \leq \delta^{\kappa '} \nu^{\frac{\kappa}{2}}\dfrac{ I_1(d^t,k)I_2(d^t) I_1(d^n,\epsilon k')}{I_1(d^n,k+\delta k')} ,
\end{equation}

where $\nu=M d^{n-t}$, $\kappa=d^{2t}-1$, and $\kappa '=d^n-1$.

\end{theorem}
The proof goes similarly to the one for Theorem~\ref{bounds}.
This bound decreases with the number of parties $n$, and  increases with  the local dimension $d$ and the locality factor $t$. 
The factor $\delta^{d^n-1}$, however, makes the bound very small. \\

\begin{figure}
\centering
\includegraphics[width=1\linewidth]{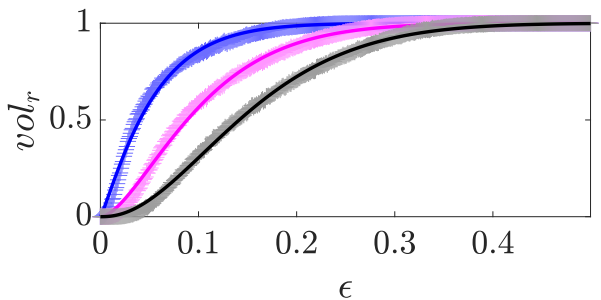}
\caption{Approximate relative volume of $\mathbf{H}'_\text{Ising}$ (with periodic boundary conditions), with $J_i\in[0,2]\; \forall i$, and $g=1$, such that the state fidelity $F(\ket{\phi},\ket{\sigma})=|\braket{\phi}{\sigma}|^2$ between its ground state and that of $H$ (for $g=J=1$) is larger or equal to $1-\epsilon$ (crosses), together with the Beta CDF that approximates it (solid lines). For $n=4$ (blue), $n=6$ (magenta) and $n=8$ (black).} 
\label{fig:plotgardnerising}
\end{figure}

\gs{\section{Numerical study: transverse-field Ising chain}}
To analyze the performance of a quantum simulator using relative volumes under a more realistic scenario,
we now take a numerical route. We consider the transverse-field quantum Ising model in 1D with Hamiltonian 
$H=\sum_{i=1}^{n} J\sigma_i^z \sigma_{i+1}^z+g\sigma^x_i$ and ground state $\ket{\psi_0}$, which can be analytically obtained using a Jordan-Wigner transformation. Here $g$ is the magnitude of the external magnetic field, and $J$ the coupling between spins. Assume now that the spin-spin interactions deviate from the constant value $J$, so that the translational symmetry is broken and the Hamiltonian reads  $H'=\sum_{i=1}^{n}  J_i\sigma_i^z \sigma_{i+1}^z+g\sigma^x_i$, with $\mathbf{H}'_{\rm Ising}$ the manifold of all such Hamiltonians. Here we estimate the probability of randomly sampling a Hamiltonian with ground state $\ket{\psi_0^\epsilon}$ from the set $\mathbf{H}'_\text{Ising}$ \gs{(we have realistically assumed that the quantum simulator at hand is only able to implement Hamiltonians from the set $\mathbf{H}'_\text{Ising}$)}. The ratio between those  and the total number of sampled $H'$ gives us an estimation of the  relative volume $\text{vol}_r(\mathbf{H}'^{\ket{\psi^\epsilon_0}}_\text{Ising}):=\text{vol}_{2^n}(\mathbf{H}'^{\ket{\psi^\epsilon_0}}_\text{Ising})/\text{vol}_{2^n}(\mathbf{H}'_\text{Ising})$. 
Such ratio is  
well approximated by a Beta cumulative distribution function, as the Beta distribution is well-suited for modeling the behavior of random variables that are limited to intervals of finite length, such as in our study where $J_i\in [0,2]$. The relative volume as a function of $\epsilon$ naturally behaves as a cumulative distribution function, as depicted in Fig. \ref{fig:plotgardnerising}. 
For small $\epsilon$, we can approximate the relative volume as  $\text{vol}_r(\mathbf{H}'^{\ket{\psi^\epsilon_0}}_\text{Ising})\approx \frac{\Gamma(\alpha+\beta)}{\alpha \Gamma(\alpha)\Gamma(\beta)}\epsilon^\alpha$, where  $\alpha \sim \text{poly}(n)$ and $\alpha,\beta >0$, making it decrease with $n$.
Interestingly, the probability of sampling the desired ground state up to a small error in this setting decreases with the number of spins, which is coherent with the observation that the TI constraint $J_i = J\,\forall i$ of the target ground state becomes more restrictive as $n$ increases.  Although this behaviour seems a priori contradictory with 
what occurs when allowing for completely general Hamiltonians, notice that the relative volume here is defined w.r.t. the volume of the manifold $\mathbf{H}'_\text{Ising}$. Had the relative volume been estimated w.r.t all possible Hamiltonians in $N=2^n$, we would have recovered the results obtained in Proposition \ref{vol_rel_prop}.
\\%

\gs{\section{Discussion}}
Measure theory is a powerful tool for tackling different aspects of Hamiltonians of which one has limited knowledge. Our work provides a novel application of this tool for the computation of volumes of parent Hamiltonians independently of their specific features. We have demonstrated that the HS measure, or any other unitarily  invariant one, is appropriate to compute relative volumes of parent Hamiltonians of a target ground state up to some error. This quantity has a direct interpretation as a minimal benchmark to the performance of quantum simulators that aim at preparing a target ground state.
We have also applied our method to the physically relevant class of local Hamiltonians, obtaining in this case an upper bound to the relative volume.
The difficulty of computing an exact volume under locality constraints 
calls for the development of more convenient techniques,
which could shed further light on the interplay between the physics of locality and the geometry of the underlying Hilbert space.\\

\gs{\section*{Acknowledgements}}
Discussions with K. \.Zyczkowski are appreciated. We acknowledge financial support from the Spanish Agencia Estatal de Investigaci\'on, PID2019-107609GB-I00, from
Secretaria d’Universitats i Recerca del Departament d’Empresa i Coneixement de la Generalitat de Catalunya, co-funded by the European Union Regional Development Fund within
the ERDF Operational Program of Catalunya
(project QuantumCat, ref. 001-P-001644), and
Generalitat de Catalunya CIRIT 2017-SGR-
1127.

\bibliographystyle{apsrev4-2}
\bibliography{bibliografia}

\begin{thebibliography}{24}%
\makeatletter
\providecommand \@ifxundefined [1]{%
 \@ifx{#1\undefined}
}%
\providecommand \@ifnum [1]{%
 \ifnum #1\expandafter \@firstoftwo
 \else \expandafter \@secondoftwo
 \fi
}%
\providecommand \@ifx [1]{%
 \ifx #1\expandafter \@firstoftwo
 \else \expandafter \@secondoftwo
 \fi
}%
\providecommand \natexlab [1]{#1}%
\providecommand \enquote  [1]{``#1''}%
\providecommand \bibnamefont  [1]{#1}%
\providecommand \bibfnamefont [1]{#1}%
\providecommand \citenamefont [1]{#1}%
\providecommand \href@noop [0]{\@secondoftwo}%
\providecommand \href [0]{\begingroup \@sanitize@url \@href}%
\providecommand \@href[1]{\@@startlink{#1}\@@href}%
\providecommand \@@href[1]{\endgroup#1\@@endlink}%
\providecommand \@sanitize@url [0]{\catcode `\\12\catcode `\$12\catcode
  `\&12\catcode `\#12\catcode `\^12\catcode `\_12\catcode `\%12\relax}%
\providecommand \@@startlink[1]{}%
\providecommand \@@endlink[0]{}%
\providecommand \url  [0]{\begingroup\@sanitize@url \@url }%
\providecommand \@url [1]{\endgroup\@href {#1}{\urlprefix }}%
\providecommand \urlprefix  [0]{URL }%
\providecommand \Eprint [0]{\href }%
\providecommand \doibase [0]{https://doi.org/}%
\providecommand \selectlanguage [0]{\@gobble}%
\providecommand \bibinfo  [0]{\@secondoftwo}%
\providecommand \bibfield  [0]{\@secondoftwo}%
\providecommand \translation [1]{[#1]}%
\providecommand \BibitemOpen [0]{}%
\providecommand \bibitemStop [0]{}%
\providecommand \bibitemNoStop [0]{.\EOS\space}%
\providecommand \EOS [0]{\spacefactor3000\relax}%
\providecommand \BibitemShut  [1]{\csname bibitem#1\endcsname}%
\let\auto@bib@innerbib\@empty
\bibitem [{\citenamefont {Dyson}(1962)}]{dyson1962}%
  \BibitemOpen
  \bibfield  {author} {\bibinfo {author} {\bibfnamefont {F.~J.}\ \bibnamefont
  {Dyson}},\ }\href {https://doi.org/10.1063/1.1703773} {\bibfield  {journal}
  {\bibinfo  {journal} {Journal of Mathematical Physics}\ }\textbf {\bibinfo
  {volume} {3}},\ \bibinfo {pages} {140} (\bibinfo {year} {1962})},\ \Eprint
  {https://arxiv.org/abs/https://doi.org/10.1063/1.1703773}
  {https://doi.org/10.1063/1.1703773} \BibitemShut {NoStop}%
\bibitem [{\citenamefont {Beenakker}(1997)}]{beenakker1997}%
  \BibitemOpen
  \bibfield  {author} {\bibinfo {author} {\bibfnamefont {C.~W.~J.}\
  \bibnamefont {Beenakker}},\ }\href
  {https://doi.org/10.1103/RevModPhys.69.731} {\bibfield  {journal} {\bibinfo
  {journal} {Rev. Mod. Phys.}\ }\textbf {\bibinfo {volume} {69}},\ \bibinfo
  {pages} {731} (\bibinfo {year} {1997})}\BibitemShut {NoStop}%
\bibitem [{\citenamefont {Frisch}\ \emph {et~al.}(2014)\citenamefont {Frisch},
  \citenamefont {Mark}, \citenamefont {Aikawa}, \citenamefont {Ferlaino},
  \citenamefont {Bohn}, \citenamefont {Makrides}, \citenamefont {Petrov},\ and\
  \citenamefont {Kotochigova}}]{Frisch2014}%
  \BibitemOpen
  \bibfield  {author} {\bibinfo {author} {\bibfnamefont {A.}~\bibnamefont
  {Frisch}}, \bibinfo {author} {\bibfnamefont {M.}~\bibnamefont {Mark}},
  \bibinfo {author} {\bibfnamefont {K.}~\bibnamefont {Aikawa}}, \bibinfo
  {author} {\bibfnamefont {F.}~\bibnamefont {Ferlaino}}, \bibinfo {author}
  {\bibfnamefont {J.~L.}\ \bibnamefont {Bohn}}, \bibinfo {author}
  {\bibfnamefont {C.}~\bibnamefont {Makrides}}, \bibinfo {author}
  {\bibfnamefont {A.}~\bibnamefont {Petrov}},\ and\ \bibinfo {author}
  {\bibfnamefont {S.}~\bibnamefont {Kotochigova}},\ }\href
  {https://doi.org/10.1038/nature13137} {\bibfield  {journal} {\bibinfo
  {journal} {Nature}\ }\textbf {\bibinfo {volume} {507}},\ \bibinfo {pages}
  {475} (\bibinfo {year} {2014})}\BibitemShut {NoStop}%
\bibitem [{\citenamefont {Gardner}\ and\ \citenamefont
  {Derrida}(1988)}]{Gardner1988}%
  \BibitemOpen
  \bibfield  {author} {\bibinfo {author} {\bibfnamefont {E.}~\bibnamefont
  {Gardner}}\ and\ \bibinfo {author} {\bibfnamefont {B.}~\bibnamefont
  {Derrida}},\ }\href {https://doi.org/10.1088/0305-4470/21/1/031} {\bibfield
  {journal} {\bibinfo  {journal} {Journal of Physics A: General Physics}\
  }\textbf {\bibinfo {volume} {21}},\ \bibinfo {pages} {271} (\bibinfo {year}
  {1988})}\BibitemShut {NoStop}%
\bibitem [{\citenamefont {\.{Z}yczkowski}\ \emph {et~al.}(1998)\citenamefont
  {\.{Z}yczkowski}, \citenamefont {Horodecki}, \citenamefont {Sanpera},\ and\
  \citenamefont {Lewenstein}}]{Zyczkowski1998}%
  \BibitemOpen
  \bibfield  {author} {\bibinfo {author} {\bibfnamefont {K.}~\bibnamefont
  {\.{Z}yczkowski}}, \bibinfo {author} {\bibfnamefont {P.}~\bibnamefont
  {Horodecki}}, \bibinfo {author} {\bibfnamefont {A.}~\bibnamefont {Sanpera}},\
  and\ \bibinfo {author} {\bibfnamefont {M.}~\bibnamefont {Lewenstein}},\
  }\href {https://doi.org/10.1103/PhysRevA.58.883} {\bibfield  {journal}
  {\bibinfo  {journal} {Phys. Rev. A}\ }\textbf {\bibinfo {volume} {58}},\
  \bibinfo {pages} {883} (\bibinfo {year} {1998})}\BibitemShut {NoStop}%
\bibitem [{\citenamefont {Bengtsson}\ and\ \citenamefont
  {\.Zyczkowski}(2006)}]{bengtsson2006}%
  \BibitemOpen
  \bibfield  {author} {\bibinfo {author} {\bibfnamefont {I.}~\bibnamefont
  {Bengtsson}}\ and\ \bibinfo {author} {\bibfnamefont {K.}~\bibnamefont
  {\.Zyczkowski}},\ }\href {https://doi.org/10.1017/CBO9780511535048} {\emph
  {\bibinfo {title} {{Geometry of Quantum States: An Introduction to Quantum
  Entanglement}}}}\ (\bibinfo  {publisher} {Cambridge University Press},\
  \bibinfo {year} {2006})\BibitemShut {NoStop}%
\bibitem [{\citenamefont {Szarek}\ \emph {et~al.}(2008)\citenamefont {Szarek},
  \citenamefont {Werner},\ and\ \citenamefont {\.{Z}yczkowski}}]{Szarek2008}%
  \BibitemOpen
  \bibfield  {author} {\bibinfo {author} {\bibfnamefont {S.~J.}\ \bibnamefont
  {Szarek}}, \bibinfo {author} {\bibfnamefont {E.}~\bibnamefont {Werner}},\
  and\ \bibinfo {author} {\bibfnamefont {K.}~\bibnamefont {\.{Z}yczkowski}},\
  }\href {https://doi.org/10.1063/1.2841325} {\bibfield  {journal} {\bibinfo
  {journal} {Journal of Mathematical Physics}\ }\textbf {\bibinfo {volume}
  {49}},\ \bibinfo {pages} {032113} (\bibinfo {year} {2008})},\ \Eprint
  {https://arxiv.org/abs/https://doi.org/10.1063/1.2841325}
  {https://doi.org/10.1063/1.2841325} \BibitemShut {NoStop}%
\bibitem [{\citenamefont {Lewenstein}\ \emph {et~al.}(2021)\citenamefont
  {Lewenstein}, \citenamefont {Gratsea}, \citenamefont {Riera-Campeny},
  \citenamefont {Aloy}, \citenamefont {Kasper},\ and\ \citenamefont
  {Sanpera}}]{Lewenstein2020}%
  \BibitemOpen
  \bibfield  {author} {\bibinfo {author} {\bibfnamefont {M.}~\bibnamefont
  {Lewenstein}}, \bibinfo {author} {\bibfnamefont {A.}~\bibnamefont {Gratsea}},
  \bibinfo {author} {\bibfnamefont {A.}~\bibnamefont {Riera-Campeny}}, \bibinfo
  {author} {\bibfnamefont {A.}~\bibnamefont {Aloy}}, \bibinfo {author}
  {\bibfnamefont {V.}~\bibnamefont {Kasper}},\ and\ \bibinfo {author}
  {\bibfnamefont {A.}~\bibnamefont {Sanpera}},\ }\href
  {https://doi.org/10.1088/2058-9565/ac070f} {\bibfield  {journal} {\bibinfo
  {journal} {Quantum Science and Technology}\ }\textbf {\bibinfo {volume}
  {6}},\ \bibinfo {pages} {045002} (\bibinfo {year} {2021})}\BibitemShut
  {NoStop}%
\bibitem [{\citenamefont {Fern{\'a}ndez-Gonz{\'a}lez}\ \emph
  {et~al.}(2015)\citenamefont {Fern{\'a}ndez-Gonz{\'a}lez}, \citenamefont
  {Schuch}, \citenamefont {Wolf}, \citenamefont {Cirac},\ and\ \citenamefont
  {P{\'e}rez-Garc{\'i}a}}]{davidPG2015}%
  \BibitemOpen
  \bibfield  {author} {\bibinfo {author} {\bibfnamefont {C.}~\bibnamefont
  {Fern{\'a}ndez-Gonz{\'a}lez}}, \bibinfo {author} {\bibfnamefont
  {N.}~\bibnamefont {Schuch}}, \bibinfo {author} {\bibfnamefont {M.~M.}\
  \bibnamefont {Wolf}}, \bibinfo {author} {\bibfnamefont {J.~I.}\ \bibnamefont
  {Cirac}},\ and\ \bibinfo {author} {\bibfnamefont {D.}~\bibnamefont
  {P{\'e}rez-Garc{\'i}a}},\ }\href@noop {} {\bibfield  {journal} {\bibinfo
  {journal} {Communications in Mathematical Physics}\ }\textbf {\bibinfo
  {volume} {333}},\ \bibinfo {pages} {299} (\bibinfo {year}
  {2015})}\BibitemShut {NoStop}%
\bibitem [{\citenamefont {Kitaev}\ \emph {et~al.}(2002)\citenamefont {Kitaev},
  \citenamefont {Shen}, \citenamefont {Vyalyi},\ and\ \citenamefont
  {Vyalyi}}]{kitaev2002}%
  \BibitemOpen
  \bibfield  {author} {\bibinfo {author} {\bibfnamefont {A.}~\bibnamefont
  {Kitaev}}, \bibinfo {author} {\bibfnamefont {A.}~\bibnamefont {Shen}},
  \bibinfo {author} {\bibfnamefont {M.}~\bibnamefont {Vyalyi}},\ and\ \bibinfo
  {author} {\bibfnamefont {M.}~\bibnamefont {Vyalyi}},\ }\href
  {https://books.google.es/books?id=08vZYhafYEAC} {\emph {\bibinfo {title}
  {Classical and Quantum Computation}}},\ Graduate studies in mathematics\
  (\bibinfo  {publisher} {American Mathematical Society},\ \bibinfo {year}
  {2002})\BibitemShut {NoStop}%
\bibitem [{\citenamefont {Kempe}\ \emph {et~al.}(2006)\citenamefont {Kempe},
  \citenamefont {Kitaev},\ and\ \citenamefont {Regev}}]{kempe2006}%
  \BibitemOpen
  \bibfield  {author} {\bibinfo {author} {\bibfnamefont {J.}~\bibnamefont
  {Kempe}}, \bibinfo {author} {\bibfnamefont {A.}~\bibnamefont {Kitaev}},\ and\
  \bibinfo {author} {\bibfnamefont {O.}~\bibnamefont {Regev}},\ }\href
  {https://doi.org/10.1137/S0097539704445226} {\bibfield  {journal} {\bibinfo
  {journal} {SIAM J. Comput.}\ }\textbf {\bibinfo {volume} {35}},\ \bibinfo
  {pages} {1070–1097} (\bibinfo {year} {2006})}\BibitemShut {NoStop}%
\bibitem [{\citenamefont {\.{Z}yczkowski}\ and\ \citenamefont
  {Sommers}(2003)}]{Zyczkowski2003}%
  \BibitemOpen
  \bibfield  {author} {\bibinfo {author} {\bibfnamefont {K.}~\bibnamefont
  {\.{Z}yczkowski}}\ and\ \bibinfo {author} {\bibfnamefont {H.~J.}\
  \bibnamefont {Sommers}},\ }\href
  {https://doi.org/10.1088/0305-4470/36/39/310} {\bibfield  {journal} {\bibinfo
   {journal} {Journal of Physics A: Mathematical and General}\ }\textbf
  {\bibinfo {volume} {36}},\ \bibinfo {pages} {10115} (\bibinfo {year}
  {2003})},\ \Eprint {https://arxiv.org/abs/0302197} {arXiv:0302197 [quant-ph]}
  \BibitemShut {NoStop}%
\bibitem [{\citenamefont {Wang}\ and\ \citenamefont
  {Schirmer}(2009)}]{wang2009}%
  \BibitemOpen
  \bibfield  {author} {\bibinfo {author} {\bibfnamefont {X.}~\bibnamefont
  {Wang}}\ and\ \bibinfo {author} {\bibfnamefont {S.~G.}\ \bibnamefont
  {Schirmer}},\ }\href {https://doi.org/10.1103/PhysRevA.79.052326} {\bibfield
  {journal} {\bibinfo  {journal} {Phys. Rev. A}\ }\textbf {\bibinfo {volume}
  {79}},\ \bibinfo {pages} {052326} (\bibinfo {year} {2009})}\BibitemShut
  {NoStop}%
\bibitem [{\citenamefont {Ozawa}(2000)}]{ozawa2000}%
  \BibitemOpen
  \bibfield  {author} {\bibinfo {author} {\bibfnamefont {M.}~\bibnamefont
  {Ozawa}},\ }\href
  {https://doi.org/https://doi.org/10.1016/S0375-9601(00)00171-7} {\bibfield
  {journal} {\bibinfo  {journal} {Physics Letters A}\ }\textbf {\bibinfo
  {volume} {268}},\ \bibinfo {pages} {158} (\bibinfo {year}
  {2000})}\BibitemShut {NoStop}%
\bibitem [{\citenamefont {Lee}\ \emph {et~al.}(2003)\citenamefont {Lee},
  \citenamefont {Kim},\ and\ \citenamefont {Brukner}}]{lee2003}%
  \BibitemOpen
  \bibfield  {author} {\bibinfo {author} {\bibfnamefont {J.}~\bibnamefont
  {Lee}}, \bibinfo {author} {\bibfnamefont {M.~S.}\ \bibnamefont {Kim}},\ and\
  \bibinfo {author} {\bibfnamefont {i.~c.~v.}\ \bibnamefont {Brukner}},\ }\href
  {https://doi.org/10.1103/PhysRevLett.91.087902} {\bibfield  {journal}
  {\bibinfo  {journal} {Phys. Rev. Lett.}\ }\textbf {\bibinfo {volume} {91}},\
  \bibinfo {pages} {087902} (\bibinfo {year} {2003})}\BibitemShut {NoStop}%
\bibitem [{\citenamefont {Pandya}\ \emph {et~al.}(2020)\citenamefont {Pandya},
  \citenamefont {Sakarya},\ and\ \citenamefont {Wie\ifmmode~\acute{s}\else
  \'{s}\fi{}niak}}]{Pandya2020}%
  \BibitemOpen
  \bibfield  {author} {\bibinfo {author} {\bibfnamefont {P.}~\bibnamefont
  {Pandya}}, \bibinfo {author} {\bibfnamefont {O.}~\bibnamefont {Sakarya}},\
  and\ \bibinfo {author} {\bibfnamefont {M.}~\bibnamefont
  {Wie\ifmmode~\acute{s}\else \'{s}\fi{}niak}},\ }\href
  {https://doi.org/10.1103/PhysRevA.102.012409} {\bibfield  {journal} {\bibinfo
   {journal} {Phys. Rev. A}\ }\textbf {\bibinfo {volume} {102}},\ \bibinfo
  {pages} {012409} (\bibinfo {year} {2020})}\BibitemShut {NoStop}%
\bibitem [{\citenamefont {Arrasmith}\ \emph {et~al.}(2019)\citenamefont
  {Arrasmith}, \citenamefont {Cincio}, \citenamefont {Sornborger},
  \citenamefont {Zurek},\ and\ \citenamefont {Coles}}]{Arrasmith2019}%
  \BibitemOpen
  \bibfield  {author} {\bibinfo {author} {\bibfnamefont {A.}~\bibnamefont
  {Arrasmith}}, \bibinfo {author} {\bibfnamefont {L.}~\bibnamefont {Cincio}},
  \bibinfo {author} {\bibfnamefont {A.~T.}\ \bibnamefont {Sornborger}},
  \bibinfo {author} {\bibfnamefont {W.~H.}\ \bibnamefont {Zurek}},\ and\
  \bibinfo {author} {\bibfnamefont {P.~J.}\ \bibnamefont {Coles}},\ }\href
  {https://doi.org/10.1038/s41467-019-11417-0} {\bibfield  {journal} {\bibinfo
  {journal} {Nature Communications}\ }\textbf {\bibinfo {volume} {10}},\
  \bibinfo {pages} {3438} (\bibinfo {year} {2019})}\BibitemShut {NoStop}%
\bibitem [{\citenamefont {Abramowitz}\ and\ \citenamefont
  {Stegun}(1965)}]{abramowitz}%
  \BibitemOpen
  \bibfield  {author} {\bibinfo {author} {\bibfnamefont {M.}~\bibnamefont
  {Abramowitz}}\ and\ \bibinfo {author} {\bibfnamefont {I.}~\bibnamefont
  {Stegun}},\ }\href {https://books.google.es/books?id=MtU8uP7XMvoC} {\emph
  {\bibinfo {title} {Handbook of Mathematical Functions: With Formulas, Graphs,
  and Mathematical Tables}}},\ Applied mathematics series\ (\bibinfo
  {publisher} {Dover Publications},\ \bibinfo {year} {1965})\BibitemShut
  {NoStop}%
\bibitem [{Note1()}]{Note1}%
  \BibitemOpen
  \bibinfo {note} {Approximating $u^\dagger P u\simeq {\protect \mbox {$1
  \protect \hspace {-1.0mm} {\protect \bf l}$}}$ is simply wrong, notice that
  for a general $N$-dimensional unitary matrix $U$ it holds $\protect
  \ensuremath {\left |\left |\protect \mathds {1}-U^\dagger P U\right |\right
  |}_{\protect \text {HS}}=\protect \ensuremath {\left |\left |U^\dagger
  (\protect \mathds {1}- P)U\right |\right |}_{\protect \text {HS}}=\protect
  \ensuremath {\left |\left |\protect \mathds {1}- P \right |\right
  |}_{\protect \text {HS}}=\protect \sqrt {N-2}\gg 0$.}\BibitemShut {Stop}%
\bibitem [{\citenamefont {Różański}\ \emph {et~al.}(2017)\citenamefont
  {Różański}, \citenamefont {Wituła},\ and\ \citenamefont
  {Hetmaniok}}]{rozanski2017}%
  \BibitemOpen
  \bibfield  {author} {\bibinfo {author} {\bibfnamefont {M.}~\bibnamefont
  {Różański}}, \bibinfo {author} {\bibfnamefont {R.}~\bibnamefont
  {Wituła}},\ and\ \bibinfo {author} {\bibfnamefont {E.}~\bibnamefont
  {Hetmaniok}},\ }\href
  {https://doi.org/https://doi.org/10.1016/j.laa.2017.07.003} {\bibfield
  {journal} {\bibinfo  {journal} {Linear Algebra and its Applications}\
  }\textbf {\bibinfo {volume} {532}},\ \bibinfo {pages} {500} (\bibinfo {year}
  {2017})}\BibitemShut {NoStop}%
\bibitem [{\citenamefont {Lee}(1997)}]{lee1997}%
  \BibitemOpen
  \bibfield  {author} {\bibinfo {author} {\bibfnamefont {J.}~\bibnamefont
  {Lee}},\ }\href {https://books.google.es/books?id=ZRQgH7FQafgC} {\emph
  {\bibinfo {title} {Riemannian Manifolds: An Introduction to Curvature}}},\
  Graduate Texts in Mathematics\ (\bibinfo  {publisher} {Springer New York},\
  \bibinfo {year} {1997})\BibitemShut {NoStop}%
\bibitem [{\citenamefont {Zhang}(2015)}]{Zhang2015}%
  \BibitemOpen
  \bibfield  {author} {\bibinfo {author} {\bibfnamefont {L.}~\bibnamefont
  {Zhang}},\ }\href {https://doi.org/10.48550/ARXIV.1509.00537} {\bibinfo
  {title} {Volumes of orthogonal groups and unitary groups}} (\bibinfo {year}
  {2015})\BibitemShut {NoStop}%
\bibitem [{\citenamefont {Hua}(1963)}]{Hua:1963:HAF}%
  \BibitemOpen
  \bibfield  {author} {\bibinfo {author} {\bibfnamefont {L.~K.}\ \bibnamefont
  {Hua}},\ }\href@noop {} {\emph {\bibinfo {title} {Harmonic Analysis of
  Functions of Several Complex Variables in the Classical Domains}}},\ \bibinfo
  {series} {Translations of Mathematical Monographs}, Vol.~\bibinfo {volume}
  {6}\ (\bibinfo  {publisher} {American Mathematical Society},\ \bibinfo
  {address} {Providence, RI},\ \bibinfo {year} {1963})\ pp.\ \bibinfo {pages}
  {iv+164},\ \bibinfo {note} {reprinted in 1979.}\BibitemShut {Stop}%
\bibitem [{\citenamefont {Wu}\ \emph {et~al.}(2021)\citenamefont {Wu},
  \citenamefont {Kondra}, \citenamefont {Rana}, \citenamefont {Scandolo},
  \citenamefont {Xiang}, \citenamefont {Li}, \citenamefont {Guo},\ and\
  \citenamefont {Streltsov}}]{wu2021}%
  \BibitemOpen
  \bibfield  {author} {\bibinfo {author} {\bibfnamefont {K.-D.}\ \bibnamefont
  {Wu}}, \bibinfo {author} {\bibfnamefont {T.~V.}\ \bibnamefont {Kondra}},
  \bibinfo {author} {\bibfnamefont {S.}~\bibnamefont {Rana}}, \bibinfo {author}
  {\bibfnamefont {C.~M.}\ \bibnamefont {Scandolo}}, \bibinfo {author}
  {\bibfnamefont {G.-Y.}\ \bibnamefont {Xiang}}, \bibinfo {author}
  {\bibfnamefont {C.-F.}\ \bibnamefont {Li}}, \bibinfo {author} {\bibfnamefont
  {G.-C.}\ \bibnamefont {Guo}},\ and\ \bibinfo {author} {\bibfnamefont
  {A.}~\bibnamefont {Streltsov}},\ }\href
  {https://doi.org/10.1103/PhysRevLett.126.090401} {\bibfield  {journal}
  {\bibinfo  {journal} {Phys. Rev. Lett.}\ }\textbf {\bibinfo {volume} {126}},\
  \bibinfo {pages} {090401} (\bibinfo {year} {2021})}\BibitemShut {NoStop}%
\end{thebibliography}%



\onecolumngrid
\appendix

\section{Proof of Propositions \ref{volume_all}, \ref{vol_fixedground} and \ref{vol_fixed_eps}}\label{appendix1}
Using $H=UDU^\dagger$ one obtains
$dH= U(dD+U^\dagger dU D-DU^\dagger dU)U^\dagger$,
which leads to
\begin{eqnarray}\label{local_samei}
ds^2&=&\sum_{i=1}^{N}(d\lambda_i)^2 +2\sum_{i<j}(\lambda_i-\lambda_j)^2|(U^\dagger dU)_{ij}|^2.
\end{eqnarray}

Now, differentiating the condition $\sum_{i=1}^{N}\lambda_i= \tilde{k}\leq k$, one gets $\sum_{i=1}^{N}d\lambda_i= 0$, which implies $d\lambda_{N}=-\sum_{i=1}^{N-1}d\lambda_i$. Then
\begin{eqnarray}
\sum_{i=1}^{N}(d\lambda_i)^2&=&\sum_{i=1}^{N-1}(d\lambda_i)^2+(\sum_{i=1}^{N-1}d\lambda_i)^2 = \sum_{i,j=1}^{N-1}d\lambda_i g^{(\lambda)}_{ij}d\lambda_j,
\end{eqnarray}
where $g^{(\lambda)}=\mathds{1}_{N-1}+J_{N-1}$, with $J_N$ an $N$-dimensional matrix of ones, is a metric tensor with determinant $\det g^{(\lambda)}=N$.

Notice that, since the two sets of variables $\{d\lambda_i\}$ (with metric tensor $g^{(\lambda)}$) and $\{\re (U^\dagger dU)_{ij},\im (U^\dagger dU)_{ij}\}$ (with metric tensor $g^{(U)}$) do not get mixed up in the line element, the global metric $g$ is block-diagonal and its determinant is given by $\det g= \det g^{(\lambda)} \det g^{(U)}=N[\prod_{i<j} 2(\Lambda_i-\Lambda_j)^2 ]^2$, which is positive since $H$ is a Riemannian manifold.

The volume element of a Riemannian manifold gains a factor $\sqrt{|\det g|}$ \cite{lee1997}. Thus,
\begin{eqnarray}
dV&=&\sqrt{N}\prod_{i=1}^{N-1}d\lambda_i \prod_{i<j} (\lambda_i-\lambda_j)^2  |\prod_{i<j} 2\re(U^\dagger dU)_{ij}\im(U^\dagger dU)_{ij}|,
\end{eqnarray}
which has the form  $dV=d\mu(\lambda_1,...,\lambda_{N})\times d\nu_{\text{Haar}}$,
where $d\mu(\lambda_1,...,\lambda_{N})$ depends only on the eigenvalues of $H$ and $\nu_{\text{Haar}}$ is the Haar measure on the complex flag manifold $Fl_{\mathbb{C}}^{(N)}:=U(N)/[U(1)^N]$. Indeed, the following invariant metric can be defined on the unitary group: $ds_U^2:=d_{\text{HS}}^2(U,U+dU)=\tr(dU dU^\dagger)=\tr(U^\dagger dU dU^\dagger U)=-\tr(U^\dagger dU)^2$, where the last equality is obtained by noting that $U^\dagger U=\mathds{1}$ implies $dU^\dagger U=-U^\dagger dU$. Then, $ds_U^2=\sum_i |(U^\dagger dU)_{ii}|^2+2\sum_{i<j} |(U^\dagger dU)_{ij}|^2$, which induces the Haar measure on $U(N)$. For unitaries with fixed diagonal, that is,  $U\in Fl_{\mathbb{C}}^{(N)}$, only the second term is retrieved, yielding the Haar measure on $Fl_{\mathbb{C}}^{(N)}$ (which is present in our volume element). The Haar measure is invariant under unitary transformations, meaning that $\nu_{\text{Haar}}(V)=\nu_{\text{Haar}}(UV)$, where $V$ is a subset of $U(N)$.  


Therefore, the volume of the manifold of $N$-dimensional (complex) non-degenerate Hamiltonians with bounded trace
$\mathbf{H}_{N,k}:=\{H\in \mathcal{B}({\cal{H}}_N) : H>0;  \tr H \leq k\}$
amounts to
\begin{equation}
\text{vol}_N(\mathbf{H}_{N,k}):= \int_{\substack{H:H>0,\\\tr H\leq k}}dV=I_1(N,k)I_2(N),
\end{equation}
where 
\begin{eqnarray}\label{I1-N-k}
I_1(N,k)&=&\frac{\sqrt{N}}{N!}  \int_{0}^{\infty} \int_{0}^{k} \delta(\sum_{j=1}^{N}\lambda_j-\tilde{k}) \prod_{i<j}(\lambda_i-\lambda_j)^2\prod_{i=1}^{N}d\lambda_i  d\tilde{k} 
= \frac{\sqrt{N}}{N!} \int_{0}^{k}\frac{\tilde{k}^{N^2-1}}{\Gamma(N^2)}\prod_{j=1}^{N}\frac{\Gamma(j+1)\Gamma(j)}{\Gamma(2)}d\tilde{k}\nonumber\\
&=& \frac{\sqrt{N}}{N!} \frac{1}{\Gamma(N^2)}\prod_{j=1}^{N}\frac{\Gamma(j+1)\Gamma(j)}{\Gamma(2)}\frac{k^{N^2}}{N^2}=\left( \frac{\sqrt{N}}{N^2! \; N!} \prod_{j=1}^{N}\Gamma(j+1)\Gamma(j)\right) k^{N^2}
\end{eqnarray}
[see Eqs.~(3.37)-(3.44) in \cite{Zhang2015} and Eqs.~(4.1)-(4.3) in \cite{Zyczkowski2003}], and

\begin{eqnarray}\label{volflag}
I_2(N)&=&\int_{Fl_{\mathbb{C}}^{(N)}}|\prod_{i<j} 2\re(U^\dagger dU)_{ij}\im(U^\dagger dU)_{ij}|=\text{vol}_N(Fl_{\mathbb{C}}^{(N)})= \frac{(2\pi)^{N(N-1)/2}}{1!2!...(N-1)!}=:\frac{(2\pi)^{N(N-1)/2}}{\xi_{N-1}}.
\end{eqnarray}

A few remarks are in order. Notice that the diagonalization transformation $H=UDU^\dagger$ needs to be unique; otherwise,  the volume of $H$ would be overestimated. For that, one first has to fix the order of the eigenvalues (in our case, $0<\lambda_1<\lambda_2<...<\lambda_{N}$), since different permutations of the vector of eigenvalues pertain to the same unitary orbit. That is why we introduce the $1/N!$ factor in $I_1(N,k)$. Second, since $H=UBDB^\dagger U^\dagger$, where $B$ is a diagonal unitary matrix, $U$ is generically determined up to the $N$ arbitrary phases present in $B$. Therefore,  $U$ is uniquely specified if $U\in Fl_{\mathbb{C}}^{(N)}$. The volume of this manifold w.r.t. the Haar measure is well-known and given by Eq.~\eqref{volflag} \cite{Zyczkowski2003,Hua:1963:HAF}.
Finally notice that the second equality in \eqref{I1-N-k} can be read as the volume of states (with unit trace) times the scaling factor $\tilde{k}^{N^2-1}$ coming from the $N^2$ eigenvalue factors and the delta factor that subtracts one unit.
\\

Now, to calculate the volume of the manifold of Hamiltonians that have a specified ground state $\ket{\psi_0}$, denoted by $\mathbf{H}_{N,k}^{\ket{\psi_0}} \subset \mathbf{H}_{N,k}$, one has to  impose that one of the columns of the unitaries over which we integrate coincides with $\ket{\psi_0}$: 

\begin{eqnarray}
	\int_{Fl_{\mathbb{C}}^{(N)}}\delta(|\bra{\psi_0}U\ket{0}|-1)|\prod_{i<j} 2\re(U^\dagger dU)_{ij}\im(U^\dagger dU)_{ij}| 
	&=&\text{vol}_{N-1}(Fl_{\mathbb{C}}^{(N-1)}) = \frac{(2\pi)^{(N-1)(N-2)/2}}{1!2!...(N-2)!} \nonumber\\
	&=&I_2(N-1),
\end{eqnarray}
where $\ket{0}=(1,0,...,0)^T$ and so $U\ket{0}$ denotes the first column of $U$.

The integration over the eigenvalues does not change, so we have
\begin{equation}
	S_N^{(1)}(\mathbf{H}_{N,k}^{\ket{\psi_0}})=I_1(N,k) I_2(N-1).
\end{equation}
Note that the volume of Hamiltonians with a target ground state is actually a hypersurface. In turn, fixing $L$ eigenstates implies $S_N^{(L)}=I_1(N,k)I_2(N-L)$.\\

If instead one wants to compute the volume of  Hamiltonians with a given ground state $\ket{\psi_0}$ up to error $\epsilon$ in overlap, one needs to  impose that one of the columns of the unitaries in $I_2$ is approximately $\ket{\psi_0}$:

\begin{eqnarray}
	&\int_{Fl_{\mathbb{C}}^{(N)}}& \mathds{1}_{[1-\epsilon,1]}(|\bra{\psi_0}U\ket{0}|)|\prod_{i<j} 2\re(U^\dagger dU)_{ij}\im(U^\dagger dU)_{ij}|   \nonumber \\
	&\approx&\int_{Fl_{\mathbb{C}}^{(N)}}\epsilon \delta(|\bra{\psi_0}U\ket{0}|-1)|\prod_{i<j} 2\re(U^\dagger dU)_{ij}\im(U^\dagger dU)_{ij}|   =\epsilon I_2(N-1),
\end{eqnarray}
with
$\ket{0}=(1,0,...,0)^T$ and $\mathds{1}_{[1-\epsilon,1]}(x)$ the indicator function being 1 for $x\in [1-\epsilon,1]$ and 0 otherwise. Note that the approximation  is valid for sufficiently small $\epsilon$.

The integral over the eigenvalues is the same, so  finally 
\begin{equation}
\text{vol}_N(\mathbf{H}_{N,k}^{\ket{\psi_0}})\approx \epsilon I_1(N,k) I_2(N-1).
\end{equation}

As a consequence, the relative volume of Hamiltonians with a target state up to error $\epsilon$ is given by

\begin{equation}
\text{vol}_r(\mathbf{H}_{N,k}^{\ket{\psi_0^\epsilon}}):=\frac{\text{vol}_N(\mathbf{H}_{N,k}^{\ket{\psi_0^\epsilon}})}{\text{vol}_N(\mathbf{H}_{N,k})}=\epsilon (2\pi)^{1-N}(N-1)!. 
\end{equation}    \\

All the previous results hold when considering complex Hamiltonians. However, it is also of interest to obtain the relative volume of the subset of real Hamiltonians: as recently argued in \cite{wu2021}, it is experimentally easier to implement real states (\textit{rebits}) and real operations in  a  single-photon  interferometer setup when compared to general states and operations. Knowing that a real $N$-dimensional Hamiltonian is diagonalized as $H=ODO^T$, where $O$ is an orthogonal matrix, suffices to extend our results to the domain of real Hamiltonians. In this case, $I_2(N)$ corresponds to the volume of the real flag manifold \cite{Zyczkowski2003}, that is
\begin{eqnarray}
I_2(N)&=&\text{vol}_N(Fl_{\mathbb{R}}^{(N)})= \frac{(2\pi)^{N(N-1)/4}\pi^{N/2}}{\Gamma(\frac{1}{2})...\;\Gamma(\frac{N}{2})},
\end{eqnarray}

implying
\begin{eqnarray}
\text{vol}_r^{(\mathbb{R})}(\mathbf{H}_{N,k}^{\ket{\psi_0^\epsilon}})&\approx& \epsilon \frac{\text{vol}_{N-1}(Fl_{\mathbb{R}}^{(N-1)}) }{\text{vol}_N(Fl_{\mathbb{R}}^{(N)})} = \epsilon \;  2^\frac{(1-N)}{2}\pi^{-\frac{N}{2}}\Gamma(\frac{N}{2}).
\end{eqnarray}

\section{Proof of Theorem \ref{bounds}}\label{appendix2}

For the sake of clarity, we first demonstrate Theorem 5
for $2$-local Hamiltonians and eventually generalize the proof to the $t$-local case.

Consider the manifold $\mathbf{H}_{N,k,2}^{\rm TI}$ of $2$-local $N$-dimensional TI Hamiltonians on a chain $H=\sum_{i=1}^M h_i$, with locally equal subhamiltonians 
$h_i\equiv \mathds{1}\otimes\cdots\otimes\mathds{1}\otimes h^{(i)}\otimes\mathds{1}\otimes\cdots\otimes\mathds{1}$,
where $h$ is a $d^2$-dimensional Hamiltonian acting on two $d$-dimensional parties, and the multi-index $i$ refers to the first particle in which $h$ acts. 
For the purpose of this proof, we do not require $H>0$, but only $\tr H\leq k\in \mathbb{R}$. Each subhamiltonian can be written as $h=\sum_{i,j=0}^{d^2-1}\alpha_{ij} \sigma_i \otimes \sigma_j$, where $\sigma_i$ are the generators of $SU(d)$ plus the identity and $\alpha_{ij}\in \mathbb{R}$. If $M=n-1$, the line element of this manifold reads
\begin{eqnarray}
ds^2&=&\tr(dH^2)=\sum_{i=1}^M \tr(dh_i^2)+\sum_{i\neq j}\tr(dh_i dh_j)\nonumber \\
&=&d^n(M\sum_{i,j=0}^{d^2-1}d\alpha_{ij}^2+2((M-1)\sum_{j=1}^{d^2-1}d\alpha_{0j}d\alpha_{j0}+\binom{M}{2}d\alpha_{00}^2)). 
\end{eqnarray}
Now, since $\tr H=M d^n \alpha_{00}=k'\leq k$, $d\alpha_{00}=0$ and the metric becomes 
$\dfrac{g}{d^n}=\bigoplus_{i=1}^{d^2-1}\begin{pmatrix}
M & M-1 \\
M-1 & M
\end{pmatrix} \oplus \bigoplus_{i=1}^{d^4-2d^2+1}M$, with determinant $\det (\dfrac{g}{d^n})= (2M-1)^{d^2-1}M^{d^4-2d^2+1}$. The volume of this manifold is then
\begin{equation}\label{volHTI2}
 \text{vol}(\mathbf{H}_{N,k,2}^{\rm TI})=\sqrt{|\det g|}\int \prod_{i,j=1}^{d^2-1}d\alpha_{ij} \int \prod_{j=1}^{d^2-1} d\alpha_{0j}d\alpha_{j0}\int_0^{\frac{k}{Md^n}} \delta(\alpha_{00}-k')d\alpha_{00}dk'\,.
\end{equation}

Consider now the previous line element without the term $\sum_{i\neq j}\tr(dh_i dh_j)$. Such line element corresponds to some manifold $\tilde{\mathbf{H}}$:
\begin{eqnarray}
d\tilde{s}^2&=&\sum_{i=1}^M \tr(dh_i^2)=
d^n(M\sum_{i,j=0}^{d^2-1}d\alpha_{ij}^2). 
\end{eqnarray}
Its metric is given by $\dfrac{\tilde{g}}{d^n}=\bigoplus_{i=1}^{d^4-1}M$, with determinant $\det(\dfrac{\tilde{g}}{d^n})=M^{d^4-1}$, yielding a volume $\text{vol}(\tilde{\mathbf{H}})$ as in Eq.~\eqref{volHTI2} with $\tilde g$ instead of $g$.
%
%
Since $\det g \leq \det \tilde{g}$, it holds that $\text{vol}(\mathbf{H}_{N,k,2}^{\rm TI})\leq \text{vol}(\tilde{\mathbf H})$. 

Note  that this argument can be extended to $t$-local TI Hamiltonians $H=\sum_{i=1}^M h_i$ in either 1D, 2D, or 3D, with $h_i\equiv h=\sum_{i,j,...,k=0}^{d^2-1}\alpha_{ij...k}\underbrace{\sigma_i \otimes... \otimes\sigma_k}_{t}$, and any value of $M$. Their associated $\tilde{g}$ metric is a diagonal matrix with repeated entry $Md^n$, whereas $g=\tilde{g}+X$, where $X$ is a matrix with vanishing diagonal. Now, since the metric is always positive definite, Hadamard's inequality \cite{rozanski2017} can be applied to show  that $\det(g)\leq \det (\tilde{g})$, implying $\text{vol}(\mathbf{H}_{N,k,t}^{\rm TI})\leq\text{vol}(\tilde{\mathbf H})$.    

In conclusion, calculating the volume associated to $d\tilde{s}^2$ will give  an upper bound for the volume of $t$-local TI Hamiltonians. In order to do so, we now impose $H>0$ and rewrite the line element as
$d\tilde{s}^2=\sum_{i=1}^{M}\sum_{k=1}^{N}(d\Lambda_{ik})^2 +\sum_{k\neq l}(\Lambda_{ik}-\Lambda_{il})^2|(dG_i)_{kl}|^2$,
where $h_i$ is diagonalized as $h_i=u_i \Lambda_i u_i^\dagger$ with $\Lambda_i=\text{diag}(\Lambda_{i1},...,\Lambda_{iN})$, $u_i$ an $N$-dimensional unitary matrix, and $dG_i=u_i^\dagger du_i$. Now, since the Hamiltonian is TI, it holds that $\Lambda_i\equiv \Lambda$ $\forall i$, where $\Lambda=\bigoplus_{i=1}^{d^{n-t}}\diag(\Lambda_1,...,\Lambda_{d^t})$, and $u_i=P_i u$ with $P_i$ a permutation matrix and $u$ an $N$-dimensional unitary. Therefore, $dG_i=u_i^\dagger du_i=u^\dagger P_i^\dagger P_i du=u^\dagger du\equiv dG$ $\forall i$. Then we have  
$d\tilde{s}^2=Md^{n-t}(\sum_{i=1}^{d^t}(d\Lambda_i)^2+\sum_{k\neq l}^{d^t}(\Lambda_k-\Lambda_l)^2|dG_{kl}|^2)$.

Finally, imposing that the trace of $h$ is fixed, i.e., $\sum_{i=1}^{d^t}d\Lambda_i=0$, one obtains $d\Lambda_{d^t}=-\sum_{i=1}^{d^t-1}d\Lambda_i$ and so

\begin{eqnarray}
d\tilde{s}^2=Md^{n-t}(\sum_{i=1}^{d^t-1}(d\Lambda_i)^2+(\sum_{i=1}^{d^t-1}d\Lambda_i)^2+\sum_{k\neq l}^{d^t}(\Lambda_k-\Lambda_l)^2|dG_{kl}|^2)=\sum_{i,j}\gamma_i q_{ij} \gamma_j,
\end{eqnarray}
with $q$ a metric tensor and $\vec{\gamma}$ the vector of integration variables. The determinant of the metric tensor is $\det(q)=\nu^{\kappa}d^t \prod_{i<j}4(\Lambda_i-\Lambda_j)^4$, where $\nu=Md^{n-t}$ and $\kappa=d^t-1+\frac{d^t!}{(d^t-2)!}=d^{2t}-1$, so the volume element gains a factor $\sqrt{|\det(q)|}$:

\begin{eqnarray}
d\tilde{V}= \nu^{\frac{\kappa}{2}} d^{\frac{t}{2}} \prod_{i<j}(\Lambda_i-\Lambda_j)^2\prod_{i=1}^{d^t-1}d\Lambda_i |\prod_{i<j}2 \re(dG_{ij})\im(dG_{ij})|.
\end{eqnarray}

Recalling that $\tr H=Md^{n-t}\tr h\leq k$ and following the integration procedure in Section \ref{proof_prop_volall}, we obtain the claimed upper bound for the volume of $t$-local TI Hamiltonians:

 \begin{eqnarray}
 \text{vol}_{d^t}(\mathbf{H}_{N,k,t}^{\rm TI})\leq \nu^{\frac{\kappa}{2}}  I_1(d^t,\frac{k}{Md^{n-t}})I_2(d^t).
 \end{eqnarray}

\end{document}